# A Reliable Target Evolved Node B Selection Scheme in LTE-Advanced Handover

[1]**Sayan Kumar Ray,** [1]**NZ Jhanjhi ,**[2]**Akbar Hossain**

[1]School of Computer Science, SCS Taylor's University, Subang Jaya 47500, Malaysia

[2]Eastern Institute of Technology, New Zealand

Sayan.ray@taylors.edu.my ; Noorzaman.jhanjhi@taylors.edu.my ; Ahossain@eit.ac.nz

**Abstract**

The problem of improving the handover performance in Long Term Evolution-Advanced (LTE-A) networks has not been fully solved yet. Traditionally, the selection of the target Evolved Node B (TeNB) in the handover procedure is based on the signal strength measurements, which may not produce a reliable handover. A reliable handover method may reduce the instances of unstable or frequent handovers that otherwise waste network resources. The signal strength measurement process is inherently time consuming as the user equipment (UE) has to measure multiple neighboring eNB (NeNB) frequencies in each measurement period. An efficient handover method is required to improve the overall performance of such systems. In this paper we propose a reliable and fast TeNB selection scheme for LTE-A handover. The proposed scheme outperforms the existing LTE-A handover methods. The improved performance is achieved by selecting the TeNB based on some three independent parameters, namely orientation matching (OM), current load (CL), and the received signal strengths. An UE essentially measures only the NeNBs shortlisted based on OM and CL; thus measurement time is reduced considerably leading to a reduction of overall handover time. The performance of the proposed scheme is validated by simulation.

**Keywords:** LTE-Advanced, Orientation matching, Received signal strength, Handover zones, Polar coordinate table

## 1. Introduction

In mobile cellular networks, handover (HO) is an essential function that switches the connections of the users' mobile devices from one cell to another while roaming. Performing reliable and fast HOs are the key requirements in minimizing service disruption and facilitating seamless roaming of users in a wireless network environment. While the fast HO has been researched quite extensively by network researchers [1-3], the reliability of HO has not been fully explored yet. A HO is said to be reliable when the user equipment is successfully transferred from the service of its present base station (i.e., eNB) in the current cell to the service of its next eNB in the target cell without any call breaks (i.e., seamless calls indicating no deterioration in signal strengths or quality of services). To improve HO reliability one can reduce the instances of unstable or frequent HOs which otherwise waste the network resources. However, selecting a wrong TeNB for HO may hamper the HO reliability and may result in further unwanted HO activities. To address the capability of supporting massive user traffic capacity and data rates in the current Fourth Generation (4G) and future Fifth Generation (5G) of networks, the world is experiencing a substantial growth in the number of small (pico and femtocells) and macro cells [4-5]. Providing efficient and seamless mobility management along with reliable and fast HO between macrocells as well as macro and femtocells is becoming significantly essential [6]. Reliable HO is also important for applications such as seamless vehicle-to-vehicle (V2V) and Vehicle-to-Infrastructure (V2I) communications. In the latter, user equipment inside a moving vehicle can connect to an on-board wireless adaptor and can access the Internet through the road-side infrastructure (LTE eNBs / WiMAX base stations) [7-8]. The concept of reliability for the network as a whole is gaining much importance with the advent of self-organizing capabilities (e.g., self-healing and self-configuration) of LTE-A [9]. However, HO reliability is not the same as network reliability and the current work focuses on the former.

Traditionally HOs are performed on signal strength measurements where TeNB is chosen on the basis of received signal strength (RSS) values of the NeNBs. However, in high speed mobility these measurements may be unreliable [10]. To realize a reliable HO a TeNB should be selected based on multiple independent parameters. The proposed TeNB selection scheme considers (a) the movement direction of the UE; (b) the current load of the UE; (c) the received signal strengths of eNBs. So, the NeNB that lies close to the UE's movement direction and is least loaded (i.e., to offer good QoS) and offers the strongest signal strength is finally chosen as TeNB. The proposed scheme conforms to the LTE-A HO requirements [11].

The deployment of Third Generation Partnership Project (3GPP) LTE-based broadband systems is progressing on a larger scale and it is anticipated that within the next few years there will be around 1.6 billion LTE users worldwide [12]. While, the current commercial deployments are mainly based on 3GPP pre-Rel 12 releases (e.g., Rel 8-10), release 12 is set for completion by 2014 with expected potential deployments by the end of 2015 [13-14] and LTE-B thereafter [15]. In LTE-A networks hard HO is used by default [16-17]. Figure 1 illustrates the UE HO mechanism in an LTE-A architecture. At each LTE HO, the entire user context including the user plane packets and control plane contexts are transferred from the SeNB to the TeNB utilizing the backhaul resources of the X2 interface connecting the eNBs [18]. TeNB is selected based on the strongest received signal strengths (RSS) from different NeNBs. On receiving the measurement configuration message (Measurement Control REQ) from the SeNB, the UE performs channel measurements (through scanning) of the different NeNBs within the pre-defined measurement period. It then sends back periodic reports to the SeNB containing the processed measurement results after filtering out the effect of fast fading and layer 1 measurement / estimation imperfections [19-21].

Based on the measurement reports, the SeNB selects the TeNB and negotiates with it through the X2 interface (if X2 is unavailable then S1 interface is used) to setup a data forwarding path between the SeNB and the TeNB. The TeNB reserves resources for UE to provide unhampered QoS after the HO [19]. This is because in LTE-A the HO process is 'backward' in nature (i.e., a backward HO). Once TeNB is prepared, the UE disconnects from the SeNB and performs the network re-entry activities to resume connection with the TeNB. Details of the LTE-A HO technique can be found in [1]. In the rest of the paper we will use the terms LTE and LTE-A interchangeably.

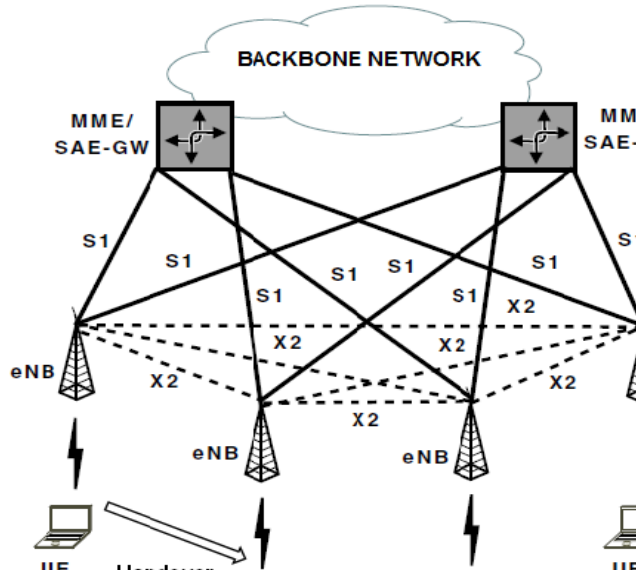

Fig.1. LTE-A Deployment Architecture

Selection of the TeNB in LTE-A HO is based on RSS measurements of the NeNBs. Different HO techniques based on the signal strength measurements are proposed for LTE and LTE-A networks [22-24]. Variables such as HO margin (HOM) and Time to Trigger (TTT) timers are introduced in LTE and LTE-A networks to reduce ping pong effects ensuring that the most appropriate TeNB is selected for the HO [16]. The HO margin is constant representing the threshold of the difference in RSS values between the serving and the target eNBs, and a TTT value is the time interval that is required for satisfying the HOM condition [20], [23-24]. However, as loads of eNBs are not considered to be a potential decisive factor in these works, ping pong effects or service disruption can still occur if the chosen TeNB is overloaded or near overloaded. A TeNB selection scheme is proposed in [25] taking into consideration both the RSS measurements and eNB loads exchanged over the X2 interfaces. However, as the eNBs do not share the load information instantly, possibility of SeNB receiving outdated load status report is high. A least-mean square (LMS)-based TeNB selection process for an LTE hard HO activity that estimates the time-varying load change of the NeNBs is proposed in [26]. Through the use of more accurate and updated eNB load information, this method showed an improvement in the LTE HO failure probability. However, most of the existing proposals did not consider the UE's movement direction when selecting the TeNB. This movement direction is very important for both the reliability and fast HO perspective. This is because predicting the UE's direction of movement well in advance may shorten the NeNB measurement phase. An UE can omit all NeNBs not lying in its movement path from scanning, which may lead to shorter measurement period and reduced HO delay. Thus the HO performance particularly for high-speed delay-sensitive applications could be improved [27]. In the HO methods described in [1], [28], the TeNB is selected based on the UE's position and future direction of motion. With reduced measurement-related delays, both the schemes have potential to generate faster HOs. However, these two schemes did not consider the load of the individual eNBs while selecting the best TeNB [43-45].

In summary, the TeNB is selected primarily based on quality of the received signal strengths, where both HO speed and reliability are compromised, or mostly based on UE's movement direction, where the reliability is compromised. None of the existing schemes reviewed in this section provide a solution both from the reliability and fast HO perspectives in LTE-A. Another limitation of traditional LTE-A hard HO is that the TeNB selection incurs HO delays and consequently fails to choose the right TeNB for a reliable HO [29]. A current LTE-compliant UE may scan up to eight frequencies in each measurement period. However, research results discussed by Nguyen et. al. [29], has indicated that even this current requirement for frequency measurement capability (of scanning up to eight frequencies) may be insufficient for choosing the appropriate TeNB and producing a reliable HO activity without considering other selection parameters. To overcome the HO problem, we propose an UE-assisted, network-controlled reliable and fast TeNB selection method for LTE-A HO based on multiple independent parameters including RSS measurements. The basic idea is that the SeNB selects the TeNB for the next HO activity based on the weighted averaging of scores for each NeNB assigned against the three different parameters, namely, the orientation matching (OM) between the geographical position of each NeNB and the UE's broad direction of motion (both with respect to the SeNB), the CL of each NeNB and the RSS-based measurements recorded by the UE from each NeNB. These multiple parameters ensure that the TeNB is rightly chosen and UE stays for maximum possible period under the TeNB (after the HO). Chances of ping pong activities are also reduced. Moreover, as the UE only measures NeNBs shortlisted based on these parameters, less time is required for measurement and the overall HO becomes faster. The scheme also employs novel ways of measuring current loads of NeNBs and assigning individual scores to each of them. Our preliminary work has been published in [30].

The rest of the paper is organized as follows. The proposed HO method is presented in Section 2. The concept of RSS-based zones for efficient HO is described in Section 3. Details of orientation matching are discussed in Section 4 and the assignments of scores to different parameters and the selection of TeNB is discussed in Section 5. Section 6 validates results by simulation. The benefits and practical implications are discussed in Section 7, and a brief conclusion in Section 8 ends the paper.

## 2. The Proposed Handover Scheme

This paper presents a novel network-controlled mechanism for reliable and fast selection of TeNB for HO in LTE-A networks. The UE plays an important role in the HO process. Recall that the proposed scheme employs three independent parameters such as OM, CL and RSS to select the TeNB. We introduce the concept of RSS power-based zones to efficiently manage the process of target eNB selection and the overall HO activity. Four conceptual zones, namely, the Zone of Normalcy (ZN), the Zone of Concern (ZC), the Zone of Emergency (ZE) and the Zone of Doom (ZD) are created by UE [3]. The zones are created by partitioning the dynamic range $[0, P_m]$ of the RSS power P, that an UE receives from its SeNB, into three different levels, $P_1$, $P_2$ & $P_3$, $P_1<P_2<P_3$. The zones correspond to RSS powers lying in the ranges $(P_m \geq P > P_3)$, $(P_3 \geq P > P_2)$, $(P_2 \geq P > P_1)$ and $(P_1 \geq P)$, respectively. The UE continuously monitors the RSS power of its SeNB to identify the zone it is currently in.

The proposed method of target eNB selection and HO is outlined below. Immediately after entering the cell of a new eNB (i.e., the new SeNB for the UE), the UE (hand held or mounted on a vehicle) sends a Status Report message to the SeNB and continues its independent motion. This message is a new LTE-A HO MAC-management message that informs the SeNB about the present direction or orientation of the UE's motion. The message contains a dynamically maintained Visited eNB List (VeNBL) that stores the chronological sequence of the eNB-Identifiers (eNB_ID) of up to K SeNBs that the UE has most recently visited. Also, the UE continuously monitors signals from the current SeNB to find out the current zone it is in. At this stage, the UE is expected to be at the ZN. On receiving this VeNBL, the current SeNB performs an OM procedure between the UE's direction of motion as represented by the VeNBL and the geo-location orientation of the centroid of each NeNB using a Polar Coordinate Table (PCT) maintained by it. Let us assume that every eNB stores in its PCT the polar coordinate of every other eNB (with respect to its own centroid as the origin of this polar coordinate system) against the latter's eNB_ID. Based on the OM performed, the SeNB assigns an OM score, $S_{OM}$ to each NeNB. Those NeNBs, whose geo-locational orientations with respect to the direction of the UE's motion represent a progressive or forward movement for the UE, are assigned a positive $S_{OM}$ while others get a negative $S_{OM}$.

When the signal strength from the current SeNB drops, i.e., P has equalled or dipped below $P_3$ ($P_3 > P > P_2$), it implies that the UE has entered the ZC. At this stage, the SeNB collects the information about the CL of each NeNB (through the X2 interfaces) and assigns individual CL-based scores, $S_{CL}$, to them. In an LTE-A network, the different eNBs can directly communicate and exchange information with each other through the X2 interface. Any overloaded NeNBs that are unlikely to be able to offer satisfactory QoS to additional connections (or may even lead to call drops) are assigned a negative $S_{CL}$. In the context of a cellular network scenario, load of eNBs changes over a time frame of minutes. In the proposed method, as the SeNB gathers the CL information almost immediately before the UE switches to the TeNB, we can justifiably argue that CL status of NeNBs remain unchanged till the completion of the HO activity. Next, the SeNB checks the two scores $S_{OM}$ & $S_{CL}$ of each NeNB, identifies any NeNB with a negative score and sends Measurement Control REQ message to the UE containing not only the parameters to measure and their thresholds but also IDs of all NeNBs having positive scores for both OM and CL. So, all those NeNBs that either have a negative score for $S_{CL}$ or a negative score for $S_{OM}$ are omitted leaving the UE to only scan few NeNBs having positive scores. The UE then scans these shortlisted NeNBs through which it records their RSS-related measurement values. The UE then sends the Measurement Report (with the recorded values) back to the SeNB when the reporting threshold conditions are fulfilled [19]. On receiving the RSS measurement values of those shortlisted NeNBs, the SeNB assigns the signal strength score $S_{RSS}$ to each of them [40-42]. From here on we will call these shortlisted NeNBs as the potential target eNBs (PTeNB). The SeNB computes the weighted average of the three individual scores $S_{OM}$, $S_{CL}$ & $S_{RSS}$ of each PTeNB and chooses the PTeNB with the highest Weighted Average Score (WAS), $S_{WAS}$, as the TeNB. Finally, when P has equalled or dipped below $P_2$ ($P_2 \geq P > P_1$), it implies that the UE has entered the ZE. The UE then carries out the HO with the selected TeNB as per the X2-based LTE-A HO procedure [11].

Figure 2 shows the flow chart of the proposed HO algorithm. Two important points needs to be noted in this context. Firstly, in an LTE-A HO framework, an UE periodically scans all available NeNB frequencies, irrespective of their positions, to gather their RSS-related measurements [1]. The TeNB for the HO is then chosen mostly based on those measurement results. To successfully measure the signal quality of NeNBs, UE needs time duration in form of measurement periods, which are only available at discrete moments [29]. Within a measurement period (e.g., 200 ms), an UE simultaneously scans frequencies of eight NeNBs [31]. However, relevant research has found that the current requirement of scanning even up to eight NeNB frequencies per measurement period is still insufficient for choosing the right TeBN for a reliable HO performance [29].

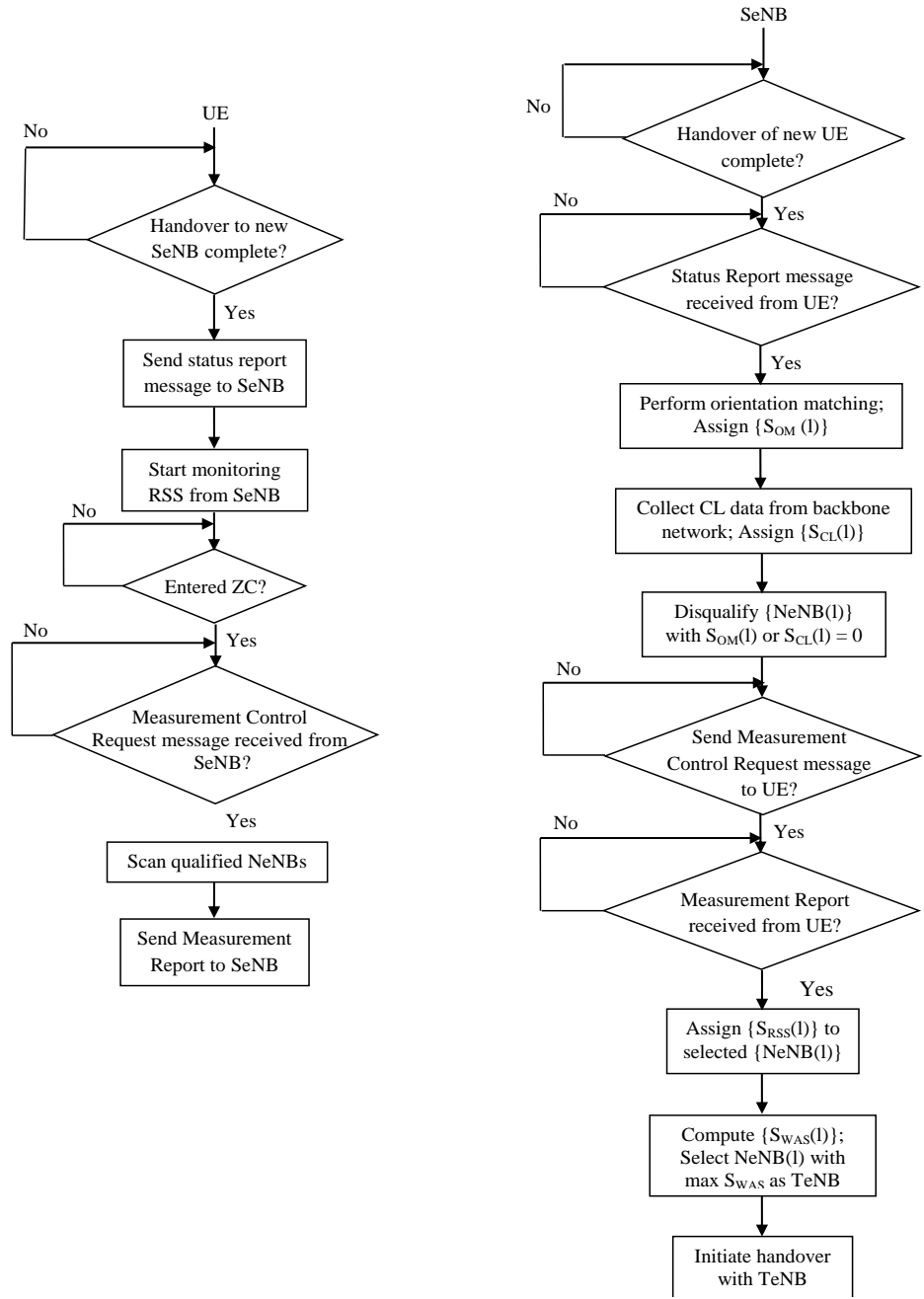

Fig.2. Flow chart of the proposed HO scheme

This is primarily because other important parameters such as UE's movement direction, load factor and QoS requirements are not considered. Secondly, the SeNB signal quality can degrade much faster than the measurement period. This may lead to a call drop or an outage in the link before a suitable TeNB is found and the HO is completed. As potential solution to these issues, as in our scheme, the TeNB is chosen based on multiple parameters (OM, CL and RSS), it provides a more reliable choice for the HO. Moreover, because an UE in our proposed scheme scans and measures less number of NeNB frequencies (only the shortlisted NeNBs), it completes the measurements much faster leading to an overall improved HO performance.

## 3. Analysis of RSS-based Zones for Efficient Handover

To efficiently manage the entire process of TeNB selection and HO, we introduce a novel concept of RSS power-based zones. By partitioning the dynamic range $[0, P_m]$ of the RSS power P, that an MS can receive from its SBS, into three different levels, $P_1$, $P_2$ & $P_3$, $P_1 < P_2 < P_3$, the MS creates four conceptual zone, namely the ZN, the ZC, the ZE and the ZD. They correspond to RSS powers lying in the ranges $(P_m \geq P > P_3)$, $(P_3 \geq P > P_2)$, $(P_2 \geq P > P_1)$ and $(P_1 \geq P)$, respectively, as shown in Figure 3. The UE periodically monitors the RSS power of its SeNB for identifying the zone it is presently in. While, in the ZN, the UE, on top of monitoring the RSS of the SeNB, also performs the OM procedure, in the ZC, the selection of the TeNB is done. Finally, the X2 HO activity takes place in the ZE. Thus, all HO-related activities are completed before the UE enters the ZD to avoid excessive packet losses or call drops which may otherwise occur owing to very poor RSS in the ZD.

From the system implementation point of view of the different zones, ZD may be defined as the zone where the RSS threatens to drop below the receiver's (i.e. UE's) sensitivity at the lowest modulation scheme (typically ½ rate QPSK). This defines the upper threshold $P_1$ of ZD that is also the lower threshold of ZE. Similarly, the lower threshold of ZN, denoted by $P_3$, may be taken to be the receiver's sensitivity at the highest modulation scheme (typically 5/6 rate 64-QAM) or one of the near highest modulation schemes to suit the operational requirements of the network operator. The lower thresholds $P_2$ of ZC and $P_1$ of ZE may be chosen to divide the interval between $P_1$ and $P_3$ into two equal parts based on the operational considerations of the network operator. Utilization of the zones in the proposed TeNB selection and HO technique is described in the next section.

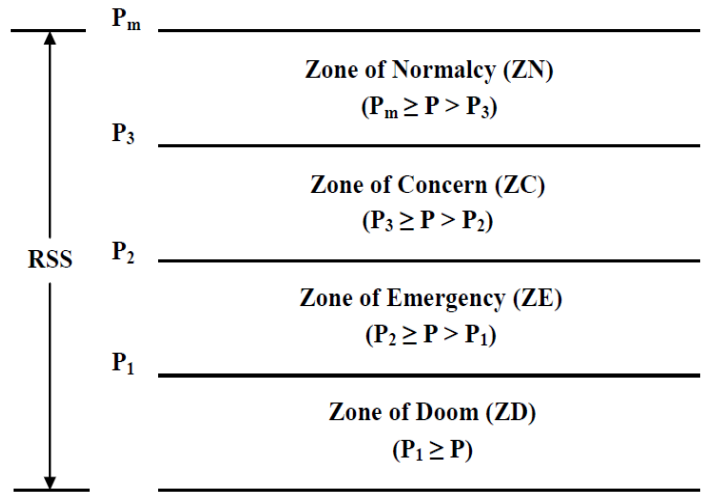

Fig.3. Zones based on RSS levels.

## 4. The Basic Concept of Orientation Matching

In this section, we describe the concepts of geo-location orientation and PCT, VeNBL and orientation matching.

### *4.1 eNBs Maintaining the Polar Coordinates of NeNBs*

Similar to base stations of all current cellular networks, LTE-A eNBs are equipped with GPS-based geo-location facilities enabling them to know the geographical position of each NeNB (with respect to an SeNB of an UE). Thus, aided by its GPS receiver, each eNB in LTE-A learns about the absolute geo-location of itself (its centroid) and communicates this information, via the backbone S1 interface, to the Packet Data Network (PDN) Gateway or PGW in the Evolved Packet Core (EPC) [11]. PGW maintains and maps the IDs of each eNB in the network to its geolocation in the form of (X, Y, Z). Thus, it is reasonable to assume that the relative positional information of the centroid of every other eNB in the network with respect to the centroid of each eNB, in polar coordinates, may either be already available in the PGW database or be computed in the PGW database without much difficulty. Alternatively, given a copy of that mapping table, each eNB can easily compute it's (with reference to its centroid as origin) own PCT [32]. Suppose A and B are two eNBs with absolute Cartesian coordinates $(x_1, y_1, z_1)$ and $(x_2, y_2, z_2)$ respectively, then the reference eNB, say A, can compute the polar coordinate $(r, \theta)$ of B with respect to its own centroid and store them in the PCT as shown in Table I (here we are neglecting the altitude Z in the X, Y, Z coordinate). Thus, we can assume that each eNB has the knowledge about the polar coordinate $(r, \theta)$ of the centroid of every other eNB in the network with respect to its own centroid.

TABLE I. PCT OF EACH ENB IN A N ENB NETWORK

| eNB_ID (i) | Polar Coordinate (j) |
|---|---|
| 1 | $r_{i1}, \theta_{i1}$ |
| 2 | $r_{i2}, \theta_{i2}$ |
| . | . |
| . | . |
| j | $r_{ij}, \theta_{ij}$ |
| j + 1 | $r_{i(j+1)}, \theta_{i(j+1)}$ |
| ... | … |
| N | $r_{iN}, \theta_{iN}$ |

*4.2 SeNB Performing Orientation Matching using UE's Direction of Motion*

We assume that in course of its whole journey, an UE dynamically maintains a VeNBL that stores the chronological sequence of the eNB_IDs of up to K SeNBs that the UE had most recently visited. Suppose, at the beginning of the UE's journey from under the SeNB S, the VeNBL is empty and all the K entries are blanks (…). Thereafter, with every HO to a new eNB, the list is updated with the eNB_ID of the new eNB appended to it and eNB_ID of the oldest (or least recent) eNB deleted from the list. Thus if the UE has just entered the coverage area of eNB M after having chronologically passed through the eNB-path S-J-K-L, the updated VeNBL entries are _ _ _ SJKLM (assuming K=8). As soon as the UE is handed over to a new SeNB, the UE passes on the VeNBL to it.

Upon receipt of the list, the $eNB_i$ (i.e. the SeNB) under consideration, goes through it and for each visited eNB $eNB_k$, k=1,2,...,k, in the list, the $eNB_i$ reads out from its PCT the stored value of the polar coordinate $(r_{ik},\theta_{ik})$ of $eNB_k$. These polar coordinates actually represent the (distance, angle) pairs of the centroids of $\{eNB_k\}$ relative to the centroid of $eNB_i$, which is imagined as the origin of $eNB_i$'s own polar coordinate system. From these k angle values or angle samples, $\{\theta_{ik}\}$, the $eNB_i$ determines the UE's "average angle of motion" (AAM) $\theta_{av}(i)$ with respect to its own polar coordinate system with the origin at its centroid. From the AAM, the $eNB_i$ can estimate the UE's "expected angle of exit" (EAE) from within its coverage area. Then, by matching the AAM and EAE with the geographical orientation (in terms of the polar coordinate stored in the PCT) of its each NeNB, the $eNB_i$ can predict, fairly well, which NeNB the UE is most likely to pass through next, provided the UE's direction of motion satisfies the reasonable assumption that during its entire long journey, the vehicle carrying the UE broadly takes nearly the shortest possible path to the destination, with no backward, random or zigzag movement, in general. However, the path may contain occasional curvatures and a few sharp bends on either side.

For determining the average angle $\theta_{av}(i)$ from the k angle samples $\{\theta_{ik}\}$, a couple of points need to be noted. Firstly, the network coverage of each eNB in the network is modeled as a circle generally having a radius in the range 500 m – 2 Km [11], [33]. Hence, during its journey from eNB1 to $eNB_i$, via the k-1 intermediate eNBs, viz., eNB2 through $eNB_k$, the UE's actual position, while it is inside the successive $eNB_k$s, could have been at any random distance $d_k$ (0 < $d_k$ < 2 km) away from the respective centroids of $\{eNB_k\}$, instead of being, ideally, on the centroids themselves. Clearly, this implies that the sequence of the k angle samples $\{\theta_{ik}\}$, k = 1, 2, ..., k, that are supposed to represent the UE's direction of motion relative to the centroid of $eNB_i$ are somewhat erroneous and the errors are random and bipolar. They are bipolar simply because, while passing through any $eNB_k$, the UE may be $d_k$ metres away (0<|$d_k$|<2 km) on either the left or the right of the centroid of $eNB_k$. It is easy to conclude that, since k >> 1, and the errors are bipolar and random, we can obtain a reasonably good estimate of the UE's angle of motion through simple averaging of the k angle values $\{\theta_{ik}\}$. However, as during the UE's journey through the successive k eNBs listed in the VeNBL, the k distance values $\{r_{ik}\}$ are not constant but reduce progressively as $r_1 > r_2 > ....... > r_k$, simple averaging will produce an incorrect result for $\theta_{av}(i)$ under this condition. This is because of the well-known trigonometric concept of "measure of an angle in radian", which is given by the relation shown in Eq. (1) below.

*Radian measure of an angle θ at the centre of a circle = (Length x of the arc of the circle that subtends the angle θ at the centre) / (Length of the radius r of the circle)* (1)

Now consider an arc of length x of a circle, with radius r and centre O, subtending an angle $\theta$ at centre O. According to Eq. (1), the parameters r, $\theta$ and x are related by Eq. (2) below.

$$r \theta = x \qquad (2)$$

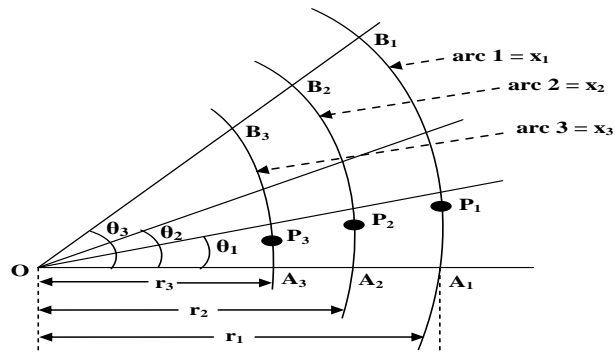

Fig. 4. Radian measure of three angles $A_1OB_1$, $A_2OB_2$ and $A_3OB_3$ subtended by arcs $A_1P_1B_1$, $A_2P_2B_2$ and $A_3P_3B_3$, respectively

To illustrate the above concept we consider Fig. 4, which shows three circles each having its centre at O and each of them has a different radii, $r_1 > r_2 > r_3$. Their respective arcs, arc1=$x_1$,

arc2=x$_2$ and arc3=x$_3$, subtend angles θ$_1$<θ$_2$<θ$_3$ at the centre. We wish to determine the average value θ$_{av}$ of the three angles. From Eq. (2), we get

$$\theta_{av} = \frac{arc_{av}}{r_{av}} = \frac{x_{av}}{r_{av}} \quad (3)$$

From Eq. (2) and in context to Figure 3, we also have

$$r_1\theta_1 = x_1 \quad (4)$$
$$r_2\theta_2 = x_2 \quad (5)$$
$$r_3\theta_3 = x_3 \quad (6)$$

By combining Eq. (3) to Eq. (6), we have

$$\theta_{av} = \{(x_1+x_2+x_3)/3\} * \{3/(r_1+r_2+r_3)\} = (r_1\theta_1+r_2\theta_2+r_3\theta_3)/(r_1+r_2+r_3) \quad (7)$$

Equation (7) shows that weighted averaging of the k angles {θ$_{ik}$} by their respective radii {r$_{ik}$} is required instead of simple averaging. Thus we can express this result more formally as

$$\text{AAM} = \theta_{av}(i) = \frac{\sum_{k=1}^{k} r_{ik}\theta_{ik}}{\sum_{k=1}^{k} r_{ik}} \quad (8)$$

## 5. Score Assignments and Selection of Target eNB

In this section, we discuss the procedure that an UE and its respective SeNB follow to select the TeNB and complete the HO activity. After receiving the VeNBL from the UE, eNB$_i$ goes through the list containing the different number of visited eNBs. Let us assume that number of previously visited eNBs is 4 and is denoted by k. eNB$_i$ then reads out from its PCT the following polar coordinates of the {BS$_k$}, k=1,2,3,4, against their respective eNB-Ids. Here all angles are in degrees.

r$_{i1}$ = 6 Km; r$_{i2}$ = 5 KM; r$_{i3}$ = 4 KM; r$_{i4}$ = 3 Km;

θ$_{i1}$ = 110°; θ$_{i2}$ = 90°; θ$_{i3}$ = 80°; θ$_{i4}$ = 100°;

Based on these values, eNB$_i$ computes the UE's AAM with respect to its centroid, using Eq. (8), as

(6*110° + 5*90° + 4*80° + 3*100°) / (6+5+4+3) = 96.11°

At this point we refer back to Section 3 and note that the "broad" direction of motion of the UE is a straight line, which makes an angle of 96.11° with the reference line at the eNB$_i$ centroid and the UE's actual path is only a "near straight line". Having estimated that the UE has entered its cell at an angle of 96.11° (approximately), eNB$_i$ infers that the UE is likely to exit the cell at an angle of (96.11°+180°)=276.11° (approximately), which is the EAE, if it continues its motion along the average direction of its journey so far for a distance of about the radius of the cell, which lies in the range 500 m – 2 Km. For a system like LTE-A, which allows the UE to move at a speed of up to 350 Km, this distance can be covered in just few seconds [11]. Here we can justifiably assume that movement directions of UEs in cellular networks do not change frequently and generally remains broadly linear over a few seconds [34]. However, just in case the UE deviates away from its direction of motion, the EAE will, of course, change from this computed value of 276.11°. For each of the NeNBs', the $eNB_i$ next read its PCT to learn about the Geographical Angle of the NeNBs (GAONB) to determine the Relative Angular Distance (RAD) between the EAE and the GAONB of its each NeNB. At this point, we arbitrarily assume that $eNB_i$ has 6 NeNBs, {$NeNB_l$}, L=1,2,...,6 and their respective GAONB (in degrees) are

GAONB 1 = 30°; GAONB 2 = 90°; GAONB 3 = 160°;
GAONB 4 = 230°; GAONB 5 = 290°; GAONB = 350°;

While computing the RAD for each of the NeNBs, $eNB_i$ takes in to account that (i) a negative sign for a RAD is meaningless and, similarly, (ii) an angle greater than 180° for RAD actually means that this apparent RAD angle value should be reduced by 180° (because of clockwise / anticlockwise interpretation) to get the actual RAD value. The computed RAD values are shown in Table II.

TABLE II. $ENB_I$ COMPUTING RAD VALUES OF NENBS

| NeNB No. | AAM | EAE (AAM + 180) | GAONB | RAD (Apparent) (EAE – GAONB) | RAD (Actual) |
|---|---|---|---|---|---|
| 1 | 96.11 | 276.11 | 30 | 246.11 | 360 – 246.11 = 113.89 |
| 2 | 96.11 | 276.11 | 90 | 186.11 | 360 – 186.11 = 173.89 |
| 3 | 96.11 | 276.11 | 160 | 116.11 | 116.11 |
| 4 | 96.11 | 276.11 | 230 | 46.11 | 46.11 |
| 5 | 96.11 | 276.11 | 290 | -13.89 | 13.89 |
| 6 | 96.11 | 276.11 | 350 | -73.89 | 73.89 |

*5.1 Score Assignment against Orientation Matching*

For assigning the scores against OM to each of the NeNBs, we chose a system of relative scoring, with only positive scores, whereby the sum of the scores of all the NeNBs, {NeNB(l)}, l=1,2,…,L, is $\sum_{l=1}^{L} S_{OM}(l) = 1$. The score $S_{OM}(l)=0$ is reserved for a "disqualified" $NeNB_l$, the reason for which is as follows. Because of the constraint $\sum_{l=1}^{L} S_{OM}(l) = 1$, the score $S_{OM}(l)=1$ is not assigned to any NeNB as that would require all the remaining NeNBs to be disqualified, i.e. have the score $S_{OM}(l)=0$. In our illustrative example, the UE, after leaving the present cell, is very unlikely to enter a cell for which the RAD of the NeNB is very large, say greater than chosen limit, which we shall call the RAD_LIMIT. A reasonable choice for the RAD_LIMIT appears to be some value which is somewhat higher than 90°.

This is because while backward movement (90°<RAD<180°) or purely random movement of the UE were considered very unlikely, side turns to left or right during the UE's journey were considered likely. So, as a reasonable choice, we choose the RAD_LIMIT as 120° and assign positive non-zero scores $S_{OM}$ to the NeNBs 1, 3, 4, 5 and 6, which have RAD≤120° and assign a zero score to NeNB 2 that has a RAD=173.89° (this RAD value indicates almost a complete backward movement for the UE). This choice of zero score is intended to disqualify NeNB 2 (or any NeNB in general) from any further consideration towards being selected as the TeNB. Thus $eNB_i$ now assigns the orientation matching score $S_{OM}$ (0<$S_{OM}$<1) to the other qualified NeNBs. It is obvious that the $S_{OM}$ assigned to a NeNB should be inversely proportional to its RAD. For instance, NeNB 5 with RAD=13.89° must receive the highest score while NeNB 3 with RAD = 116.11° must receive the lowest score. However, neither score 0 nor score 1 can be assigned, as explained before. The way $eNB_i$ assigns OM scores is as follows. The complement value of each RAD, the RAD_COMPL is considered. The individual scores are assigned as the ratio of the respective RAD_COMPL values to the sum of all RAD_COMPL values excepting those of the disqualified NeNBs. For this score assignment purpose, we also considered an appropriate Reference RAD value, called the RAD_REF, used for complementing the RAD values. Since, RAD = 0° must receive the highest possible value less than 1 and a RAD=120° must receive the lowest possible value greater than 0, we choose the RAD_REF only a little higher than 120°, say RAD_REF = 125°. With the above choice we assign the scores as computed in Table III.

TABLE III. $ENB_i$ ASSIGNING OM SCORES TO NENBS

| RAD_LIMIT = 120; | | | RAD_REF = 125; | | |
|---|---|---|---|---|---|
| NeNB | RAD | Qualified | RAD_COMPL (125 – RAD) | Sum of RAD_COMPL | Score ($S_{OM}$) |
| 1 | 113.89 | Y | 11.11 | 261.11 | 11.11 / 261.11= 0.043 |
| 2 | 173.89 | N | - | - | - |
| 3 | 116.11 | Y | 8.89 | 261.11 | 8.89 / 261.11 = 0.034 |
| 4 | 46.11 | Y | 78.89 | 261.11 | 78.89 / 261.11 = 0.302 |
| 5 | 13.89 | Y | 111.11 | 261.11 | 111.11 / 261.11 = 0.426 |
| 6 | 73.89 | Y | 51.11 | 261.11 | 51.11 / 261.11 = 0.196 |

*5.2 Score Assignment against Current Load*

To measure the CL of each NeNB, we use a technique which is simple to measure and offers a fairly static estimate of the CL. It estimates the CL by counting the number of connections currently being handled (or passing through) by an $eNB_i$. We assume that all eNBs in the network are identical in design and the maximum number of connections that can be maintained or sustained by each eNB, i.e. the connection capacity of each eNB, is N.

Next we assume that during a HO, the SeNB has L NeNBs {$NeNB_l$}, l=1,2,...,L and that the number of connections passing through the $NeNB_l$ is $M_l$, so that the $NeNB_l$ has a CL of $CL_l = M_l/N$. It is obvious that higher the value of $CL_l$, more is the CL of $NeNB_l$ and lower should be the score $S_{CL}(l)$ assigned to $NeNB_l$. To prevent any overloaded NeNB from getting selected as the TeNB and then offer poor QoS, we choose to set a higher limit CL_LIMIT of, say, 0.9, to disqualify any NeNB with CL ≥ 0.9 from being further considered for possible selection as a TeNB. We assign a score of $S_{CL}(l) = 0$ to such excessively overloaded NeNBs. To each of the remaining (tentatively) qualified NeNBs, {NeNB(l)}, we assign scores {$S_{CL}(l)$}, which are inversely proportional to their respective CLs {$CL_l$}. In this context, it should be pointed out that any of these remaining tentatively qualified NeNBs may ultimately fail to qualify as a PTeNB, if disqualified against one or both of the other two criteria toward TeNB selection, namely OM and RSS. The method of assignment of scores {$S_{CL}(l)$} to {NeNB(l)} is described here.

To assign scores to the tentatively qualified NeNBs, we first take the complement value of each $CL_l$ and call this the CL_COMPL(l). Then we assign the individual scores as the ratio of the CL_COMPL(l) values to the sum of the CL_COMPL(l) values of all the L NeNBs except the disqualified NeNBs. For computing the CL_COMPL of all the NeNBs, we choose a reference CL value CL_REF = 0.89 (since CL ≥ 0.9 indicates an overloaded NeNB) so that the CL_COMPL values {CL_COMPL(l)} of {$NeNB_l$} may be computed for each l as

$$CL\_COMPL(l) = CL\_REF - CL_l = 0.89 - CL_l \quad (9)$$

It should be noted that the CL_COMPL values of the qualified NeNBs may range between 0 – 0.89. Now, the scores for the {$NeNB_l$} will be computed as

$$S_{CL}(l) = CL\_COMPL(l) / \sum_{l=1}^{L} CL\_COMPL(l) \quad (10)$$

For the considered illustrative example, we assume that the connection capacity of each of the 6 NeNBs {$NeNB_l$}, l = 1, 2,…,6, is 500 and the present number of connections sustained, respectively, by them are {300, 250, 452, 200, 350, 150} so that their CLs are {$CL_l$} = {0.6, 0.5, 0.904, 0.4, 0.66, 0.3}. Clearly, NeNB 3 being excessively loaded (CL3 ≥ 0.9), is assigned a score of 0 and is thus disqualified from further consideration. Moreover, NeNB 2 was earlier disqualified in OM. So, the remaining 4 NeNBs, viz., NeNB 1, NeNB 4, NeNB 5 and NeNB 6, are assigned scores in proportion to their respective CL_COMPL values as shown in Table IV. We note that the sum of the 4 CL_COMPL values of NeNB 1, NeNB 4, NeNB5 and NeNB 6 equals (0.29 + 0.49 + 0.23 + 0.59) = 1.60.

*5.3 Score Assignment against Received Signal Strength Values*

To illustrate the score assignment against RSS values using the same illustrative example as discussed above for OM and CL, we assume that depending on the present distance of the UE from each NeNB (at the time of the UE's measuring of the four non-disqualified NeNBs 1, 4, 5 and 6), the RSS (in dB) received by the UE from them are: 50, 90, 60 and 30, respectively. It is obvious that the RSS score assigned to each NeNB is directly proportional to the respective RSS values. Accordingly, the scores for the four NeNBs may be computed using Eq. (11) shown below. The computed values of $\{S_{RSS}(l)\}$ of $\{NeNB(l)\}$ are shown in Table V.

TABLE IV. COMPUTATION OF CURRENT LOAD SCORE ($S_{CL}$)

| CL_LIMIT = 0.9; | | CL_REF = 0.89; | | |
|---|---|---|---|---|
| NBS No. | CL | Qualified | CL_COMPL | $S_{CL}$ |
| 1 | 0.6 | Y | 0.29 | 0.181 |
| 2 | 0.5 | N (OM) | - | - |
| 3 | 0.904 | N (CL) | - | - |
| 4 | 0.4 | Y | 0.49 | 0.306 |
| 5 | 0.66 | Y | 0.23 | 0.144 |
| 6 | 0.3 | Y | 0.59 | 0.368 |

$$S_{RSS}(l) = \frac{RSS_l}{\sum_{l=1}^{4} RSS_l} \tag{11}$$

*5.4 Weighted Averaging of Scores towards TeNB Selection*

As per the illustrative example, once the SeNB has obtained the scores of the four NeNBs (i.e., the PTeNBs) of the UE against each of the three parameters (OM, CL and RSS), it finally computes the weighted average of these scores for each PTeNB. The PTeNB that receives the highest WAS is then selected as the TeNB. The $S_{WAS}(l)$, for the PTeNBs, where l = 1, 4, 5, 6, is computed using the Eq. (12) as shown below.

$$S_{WAS}(l) = S_{OM}(l) * W_{OM} + S_{CL}(l) * W_{CL} + S_{RSS}(l) * W_{RSS} \tag{12}$$

where $W_{OM}$, $W_{CL}$ and $W_{RSS}$ are the weights, $0 \leq W_{OM}, W_{CL}, W_{RSS} \leq 1$, assigned to the three parameters, respectively, with the condition given by Eq. 13.

$$W_{OM} + W_{CL} + W_{RSS} = 1 \tag{13}$$

OM being the most important parameter out of the three, $W_{OM}$ is assigned a higher weight (0.5) compared to $W_{CL}$ and $W_{RSS}$ (both assigned 0.25). Using Equation (12), the WAS for the PTeNBs are computed and results are given in Table VI. PTeNB 4 with the highest $S_{WAS}$ is chosen by the SeNB as the TeNB for the HO activity, which is carried out immediately following the LTE-A HO procedure mentioned in [11].

TABLE V. COMPUTED VALUES OF $\{S_{RSS}(L)\}$ FOR THE MEASURED NENBS

| NeNB$_l$ | RSS$_l$ (in dB) | S$_{RSS}$(l) |
|---|---|---|
| NeNB 1 | 50 | 0.22 |
| NeNB 2 (not scanned) | - | - |
| NeNB 3 (not scanned) | - | - |
| NeNB 4 | 90 | 0.39 |
| NeNB 5 | 60 | 0.26 |
| NeNB 6 | 30 | 0.13 |

TABLE VI. $S_{WAS}$ COMPUTATION FOR PTENBS

| PTeNB$_l$ | S$_{OM}$(l) | W$_{OM}$ | S$_{CL}$(l) | W$_{CL}$ | S$_{RSS}$(l) | W$_{RSS}$ | S$_{WAS}$(l) |
|---|---|---|---|---|---|---|---|
| 1 | 0.043 | 0.5 | 0.181 | 0.25 | 0.22 | 0.25 | 0.122 |
| 4 | 0.302 | 0.5 | 0.306 | 0.25 | 0.39 | 0.25 | 0.325 |
| 5 | 0.426 | 0.5 | 0.144 | 0.25 | 0.26 | 0.25 | 0.314 |
| 6 | 0.196 | 0.5 | 0.368 | 0.25 | 0.13 | 0.25 | 0.223 |

## 6. Results and Validation of the Proposed Handover Scheme

We have developed a Python-based simulator to validate the proposed HO scheme. In the simulation topology, 400 cells are considered in a 20 x 20 square array, with each cell having one eNB in it marked by a small "cross" (x) in Figure 4. eNBs are thus arranged in a square grid format. All the eNBs are assumed to be connected to the backbone network. The terrain area considered for simulation is 20 x inter-eNB distance. We also assume that each individual eNB is aware of its location in the terrain. The vertical and horizontal spacing between two adjacent eNBs is equal to the inter-eNB distance and the range of coverage of each eNB is considered to be 75% of the inter-eNB distance. We assume that each eNB has eight NeNBs around it. We also arbitrarily assume that the distance between two grid lines is 10 m and the UE moves with a 10 m resolution. Thus the terrain may be considered as a square grid with side lengths of 10 x 20 x inter-eNB distance. There exists coverage overlap between adjacent eNBs. We assume that each eNB has a random and dynamically changing CL, lying between 0 and 1. During a HO, the current SeNB hands over the UE to the selected TeNB, which then becomes the next SeNB in the UE's movement path. We primarily aim to validate the reliability of the proposed scheme, i.e., whether the scheme result in the right choice of TeNBs for HO activities based on the three

HO parameters. We considered simulating the movements of UEs in the different situations where the user is moving (i) along the motorways or the state highways with the roads being relatively straight and not zigzag or random, (ii) in the cities with the roads/movements being straight/curvy/zigzag but not random and (iii) along the city centre having roads laid out in the form of grids. The simulations are performed in two different phases. In the first phase, five different movement paths of the UE, paths 1 through 5 (shown in red in Figure 5), are considered for running the simulation program. Each path passes through a large number of eNBs. In the second phase, we have considered three commonly used mobility models, namely, Random Waypoint [35], Random Direction [36] and Manhattan models [37] for the simulation. The UE randomly moves through different paths in accordance to the mobility model and performs multiple HOs. We assumed that while moving through the terrain, at each step, the UE performs a connectivity check with its SeNB to track the zone it's currently in. For the simulations in phase 1 we have considered three previous SeNBs in the VeNBL (i.e., a VeNBL of length three), whereas in phase 2, VeNBL lengths of 1 to 5 are considered. This implies that the method of OM is performed using three previously visited SeNBs for phase 1 and using one to five SeNBs for phase 2. The Walfisch-Ikegami model [38] is implemented to simulate the path loss behaviours. To calculate the $S_{WAS}$, for each NeNB, we have considered the following weights: 0.5 for OM and 0.25 each for CL and RSS. We primarily aim to validate the reliability of the proposed scheme, i.e., whether the scheme is resulting in the right choice of TeNBs for HO activities based on the three HO parameters. To do that, for every path of the UE, we have tracked its movement carrying out multiple successive HOs with different PTeNBs on the route (these PTeNBs after HO become the successive SeNBs for the UE). To compute the individual scores of NeNBs per HO, the simulation works as per the illustrative example explained in Section 5. We have also recorded whether the eNBs with which HOs are actually performed, match the eNBs as per the prediction of our scheme (in that case we call it a 'correct' or reliable HO) or not (an 'incorrect' HO). All the graphs presented depict results based on the method of multiple independent replications each of which continued until either the UE stopped its movement at the end of each path (for phase 1) or the time frame for the run expired (for phase 2).

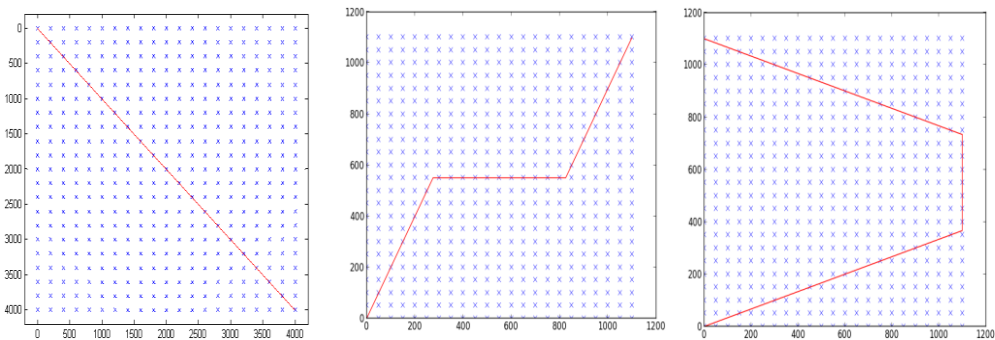

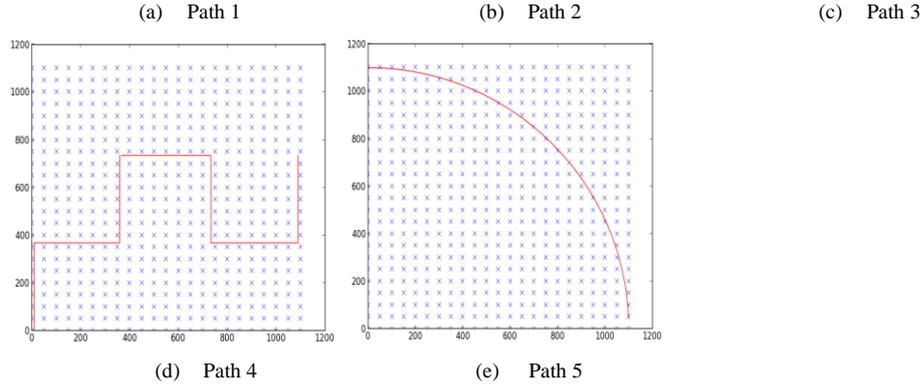

(a) Path 1  (b) Path 2  (c) Path 3
(d) Path 4  (e) Path 5

Fig 5.   UE's movement paths in the simulation topology

## *6.1 Phase 1- Results*

Figures 6 and 7 show the percentage of correct HO results for our proposed scheme in comparison to the LTE-A's conventional signal strength-based HO scheme for each movement path. In this case, we assumed NeNBs having a CL ≥ 70% as overloaded and are those are not considered as PTeNBs. Clearly, for all the paths, number of correct or reliable HOs performed by an UE as per our proposed HO scheme is much more than the conventional scheme. Results for both the proposed and conventional schemes are best for path 1 in which the UE moves in a straight path. Figure 7 shows the percentage of correct HO results when three different CL thresholds (≥ 90%, ≥ 80% and ≥ 70%) are considered. Percentage of correct HOs for ≥ 70% CL threshold is best and that for ≥ 90% is the least. The weightage considered for CL and RSS is 0.25 each and for OM its 0.5. For the proposed scheme, the fluctuations in the percentage of correct HOs are due to (i) the randomness in the assigned CL values to each NeNBs and (ii) the topology of the UE's movement paths. Owing to randomness in the CL values any NeNB that have scored well in OM and RSS, may get a poor load value and thus a low $S_{WAS}$ and misses out on the chance of getting selected as the TeNB. The reverse situation may also occur if the NeNB gets too high a score for CL and gets selected as the TeNB simply because of this high $S_{CL}$. However, implementing this technique in networks with real load numbers is expected to improve the overall reliability of the TeNB selection and HOs performed. The topology of the UE's movement path also plays a role in incorrect HO. For example, percentages of correct HOs are less for paths 2-4 owing to their sudden and sharp turns.

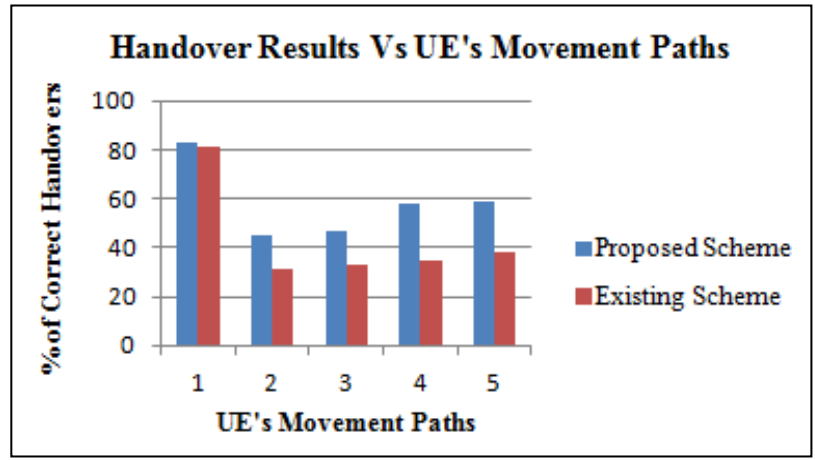

Fig.6.   Comparison of the proposed and existing LTE-A handover schemes

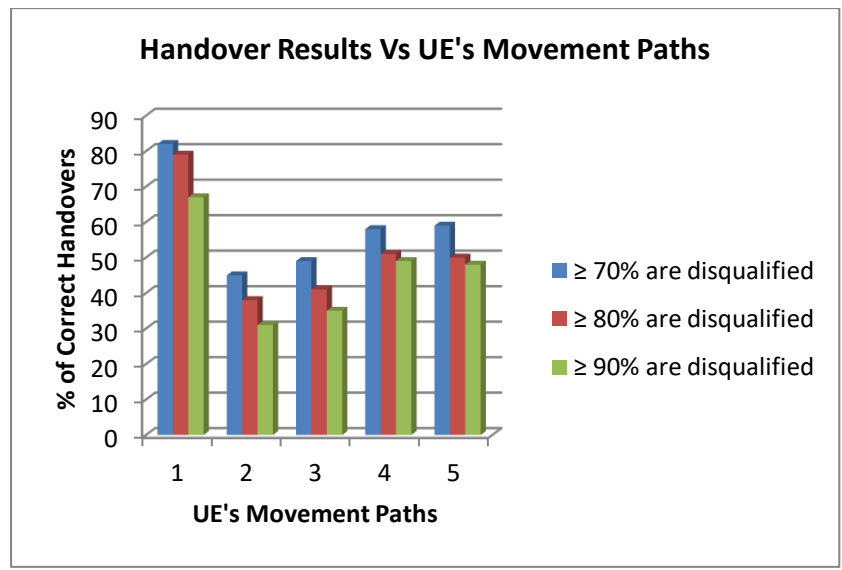

Fig. 7.   Handover results for different cut-off load thresholds

*6.2 Phase 2-Results*

In phase 2, we have considered the Random Waypoint, the Random Direction and the Manhattan models to simulate more realistic movements of the UE. We briefly, mention here the reasons for choosing these three models for the simulations. The Random Waypoint model allowed us to simulate UE's movement in different directions, but over short stretches, in a city or in the outskirts. The Random Direction model was useful to simulate UE's movements over a long stretch of path without changing directions frequently (e.g. movement along a geographical area containing a mix of relatively straight motorways and other not so straight roads). Lastly, the Manhattan model was used to simulate user movement patterns in the central part of a city, where the streets are mostly laid out in the form of grids. For the simulations in phase 2 we also considered two different eNB placement scenarios in the terrain, one with fixed inter-eNB distance and the other with a 5% variation in the inter-eNB distance. One path file is generated randomly for each mobility model. The path file produces 10000 steps for the UE to move along. Since the resolution of the field is 10m, for each step of movement the UE can move either 0m or 10m. For the Manhattan Model, the waypoints are created randomly with a resolution of 1km. The UE moves first in the x direction and then in the y direction to reach the waypoint. Upon reaching a waypoint, the UE selects another waypoint. For the Random Direction Model a random direction is chosen and the UE keeps moving until it reaches the edge of the field, at which point it will select another direction to move. Unlike the Manhattan Model, the UE can move in both x and y directions at each step, but are still restricted to move either 0m or 10m in each direction. For the Random Waypoint Model, the waypoints are again created randomly. Unlike the Manhattan Model, the resolution of the waypoints are only restricted to the fields resolution (10m). Upon reaching a waypoint, another waypoint is selected. Like the Random Direction Model, the mobile station can move in both x and y directions at each step, while still being restricted to moving 0m or 10m in each direction.

Figure 8 depicts the percentage of correct HO results for the three mobility models for a 70% cut off load threshold and 5% inter-eNB placement variation. The Manhattan model produced the most number of correct HO results (maximum of 52% when considering a VeNBL length = 3) and the Random Waypoint the least number of correct HO results (maximum of 38% when considering a VeNBL length = 3).

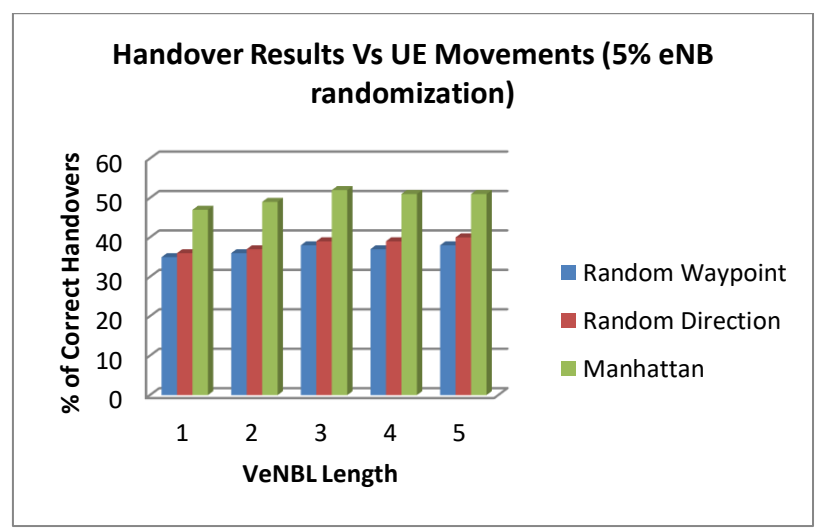

Fig. 8.  Handover results with 70% cut off load values

The Manhattan model produced the highest HO success rate and this is true for all the different simulations performed in Phase 2. This is due to the square grid like nature of the Manhattan Model corresponding to the field also being arranged in a square grid. A better prediction is made for motion in 45 degree intervals than any other angle as the grids of base stations naturally form these angles themselves. To elaborate, while choosing the next random direction to move, in case of the Random Waypoint model, the UE has to randomly choose one direction out of any direction. This is because the current movement of the UE is not dependent on the previous movement implying that every direction an UE chooses is independent of the previously chosen direction and the UE could choose any direction randomly. So every time the UE pauses to select a different movement direction, it has to choose one from any direction and this takes time. On the other hand, in case of the Manhattan model, where the roads are in the form of grids, the UE just has to choose one random direction out of only four different directions available to choose from. So, the time taken to make each of these choices is shorter than the Random Waypoint one.

We also observe that considering three previous SeNBs in the VeNBL (i.e., VeNBL length = 3) produces better success rate compared to other VeNBL values and this is also true for all Phase 2 simulation results. In Figure 7, the success rates are 52%, 38% and 39%, for the Manhattan, Random Waypoint and Random Direction models, respectively, for VeNBL length = 3. Also, according to the results the jump in HO success rate from VeNBL length = 1 to VeNBL length = 3 PBS is more compared to the success rates for VeNBL length = 3 to VeNBL length = 5. This implies that, in the current method, a VeNBL length > 3 does not generally contribute to the successful prediction of the next SeNB. Use of more number of previously visited SeNBs is susceptible to sharp turns in the movement trajectories of the UE. This is because the algorithm factors in for all the previously visited SeNBs before the sharp turn for a longer period than having few previously visited SeNBs. Using a bigger VeNBL length value is only beneficial when the UE is traveling in a straight line or a gentle curve (where the trajectory does not have sharp turns).

Figure 9 shows the results for a 70% cut off load but with fixed inter-eNB placement. Compared to 5% inter-eNB placement variations, the percentage of correct HOs is always higher in these cases and true for all other results. Manhattan model produces much better HO success rates (maximum of 75%) than the other models (maximum of 49% and 50% for Random Waypoint and Random Direction models respectively) for reasons explained above. Also, success rate is more for VeNBL length = 3 in comparison to other lengths. Here, however, we can see that for Random Direction model, VeNBL length = 4 and 5 produces slightly better success rates (almost 50%) than VeNBL = 3 (48%) and this is because of the random movement trajectory followed by the UE for this particular case.

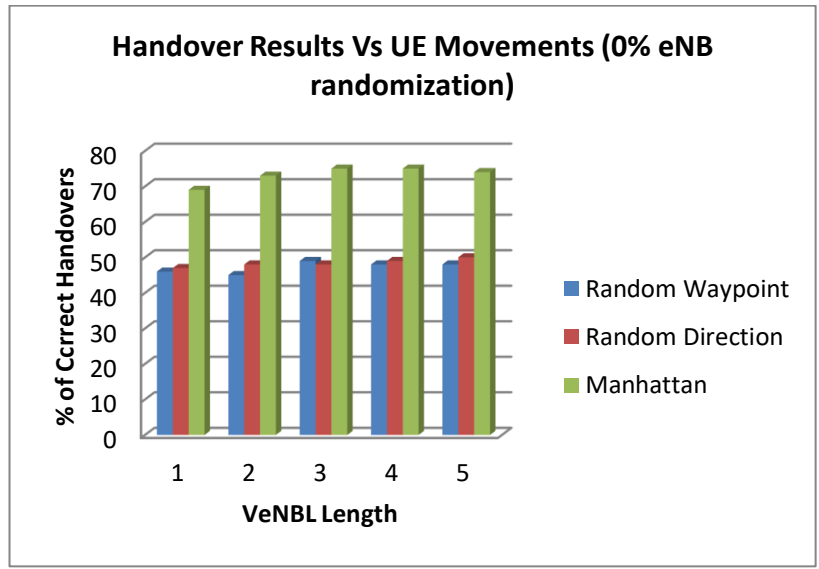

Fig. 9. Handover results with 70% cut off load values

Similarly, Figures 10 – 13 depict the number of correct HO results against VeNBL lengths 1-5 for 5% inter-eNB placement variation and fixed inter-eNB placement for 80% and 90% cut off load values. The results follow patterns similar to those discussed under figures 8 and 9. For results with 80% cut off load values (Figures 10 and 11), Manhattan model produces the maximum percentage of correct HO results (62% for fixed inter-eNB placement and 48% for eNB placement with 5% variation) and Random waypoint the minimum percentage (45% for fixed inter-eNB placement and 35% for eNB placements with 5% variation). The maximum and minimum success rates for Random Direction are 46% (fixed inter-eNB placement) and 37% (5% variation). On the other hand, as shown in figures 12 and 13, for 90% cut of load, the success percentage of correct HO is 58% and 34%, respectively, for Manhattan model with fixed inter-eNB placement and placement with 5% variation, and 27% and 20%, respectively, for Random Waypoint model with fixed eNB placement and placement with 5% variation. Similarly, for random direction, the success percentages are 32 and 29, respectively, for models with fixed eNB placement and with 5% placement variation. So, overall, percentage of correct HO results is best for 70% cut off load and fixed inter-eNB placement and least for 90% cut off load with 5% variation in inter-eNB placements, which also implies that the inter-eNB placement randomness affects the HO success rate.

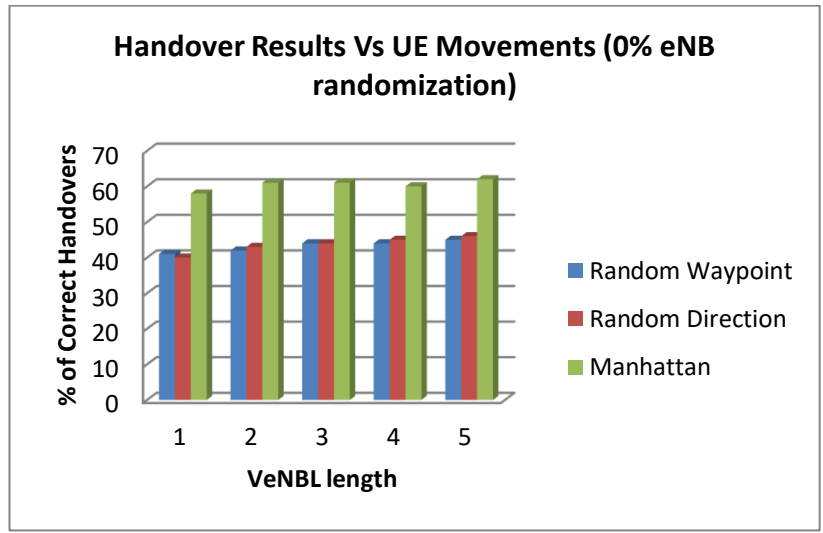

Fig.10. Handover results with 80% cut off load values

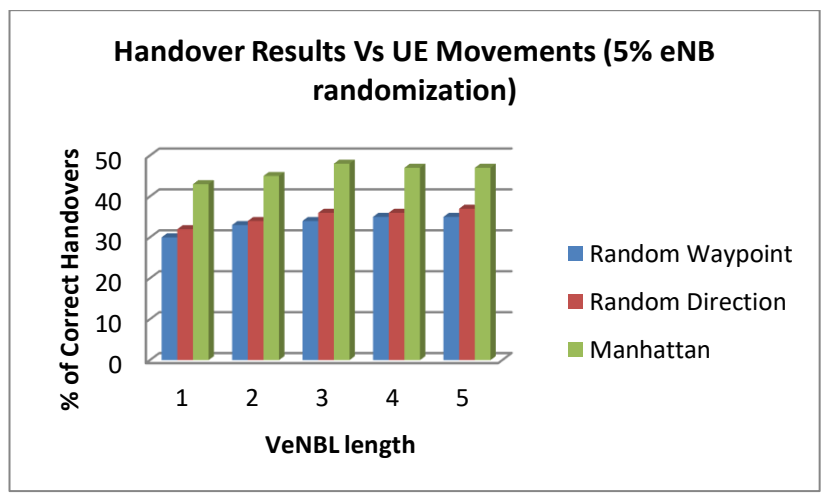

Fig.11. Handover results with 80% cut off load values

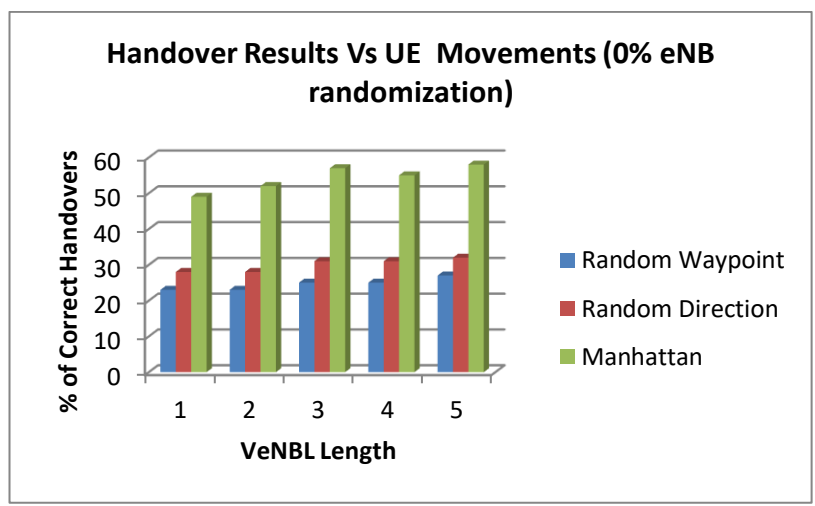

Fig.12. Handover results with 90% cut off load values

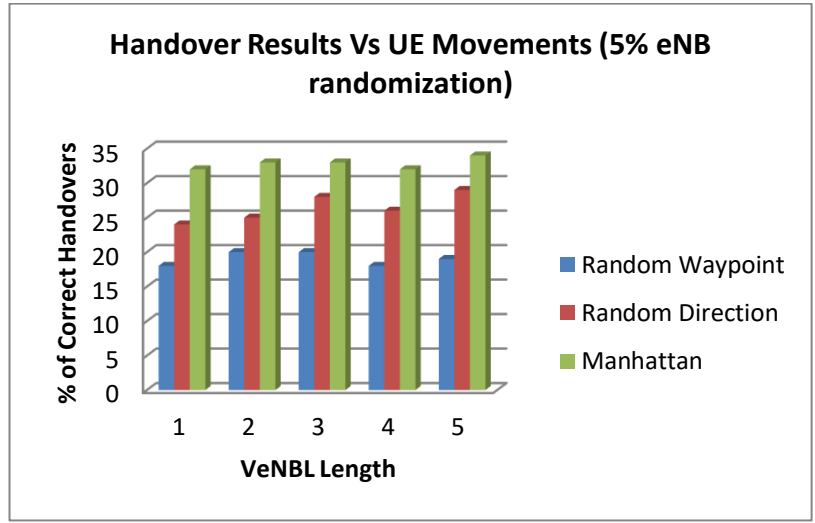

Fig.13. Handover results with 90% cut off load values

## 7. Benefits and Practical Implications

The proposed HO method offers the following three main benefits.

(i) The system provides a reliable selection of TeNB and a reliable HO performance. This benefit comes as a result of TeNB selection based on three independent parameters such as OM, CL and the RSS. In contrast, TeNB is selected one measurement values from NeNBs in the traditional HO framework.

(ii) The system offers better call dropping performance (i.e., almost zero call dropping) than the traditional HO method. The receiving inferior QoS after HO is much less indicates better system performance. This improved performance is achieved due to the characteristics of the proposed scheme where NeNBs are not chosen to be TeNB in high load conditions. The loads of NeNBs are not directly considered as a critical HO parameter in traditional LTE-A handover.

(iii) The proposed method offers faster HO than the existing HO scheme. This is because UE performs measurement activities only with the shortlisted NeNBs (based on $S_{OM}$ and $S_{CL}$). Consequently, less time is required for RSS measurements, which shortens the overall HO time. In contrast an UE performs measurement activities with all the NeNBs requiring longer time in the traditional LTE-A HO approach.

From the HO reliability perspective, the proposed system eliminates an important trade-off between an UE's measurement capability and implementation complexity (cost). Measurement capability indicates the number of cells that the UE can scan per measurement period [29]. As discussed in Section 1, a LTE-compliant terminal can scan up to eight intra-frequency cells for a given period. Although the higher measurement capability allows better HO performance but this may require more advanced and costly chipsets and more resources. In contrast, the proposed scheme offers reliable HO with reduced measurement activities as TeNB is chosen based on three key parameters (orientation matching, current load and received signal strengths). The proposed method eliminates the extremely overloaded NeNB (CL very large) in selecting TeNB ensuring a good QoS. The better QoS is achieved by avoiding possible call drops and consequently enhance HO reliability.

Moreover, during the WAS computation, $S_{RSS}$ offers a useful correction of a possible non-negligible error in $S_{OM}$. There is a possibility that after entering the current cell, the UE may unexpectedly deviate significantly from its EAE (probably by taking a left or a right turn) during the course of its journey within the cell. This will result the concerned NeNB to receive the highest $S_{OM}$, although it may not deserve it. However, the concerned NeNB will now receive a much reduced RSS score $S_{RSS}$ compared to what it would have received if the UE had not deviated considerably from its EAE. This will help to correct or neutralize the gross error margin in $S_{OM}$ to a good extent as the overall actual weighted average score $S_{WAS}$ of the NeNB will also get reduced. This good feature of the WAS scheme for TeNB selection considerably increases the reliability of the HO. However, the reliability of the proposed scheme can further be enhanced by taking two or more sets of RSS measurements (instead of just one) in quick succession and averaging the set of RSS values. However, this will increase the HO latency, though only marginally. Obviously, this will be a desirable trade off.

From the fast HO perspective, an UE initiates the process of OM immediately after entering the cellular region of its new SeNB. OM completes within a fraction of a second and thus practically introduces no delay at all. The reason for the fast execution of the process are elementary operations like memory read, add, subtract, compare, swap etc., with only K multiplications and $< (L+1)$ divisions, which are somewhat time consuming operations. The total OM process, including score assignment, is unlikely to take more than a fraction of a second even on a slow computer. Moreover, the proposed method introduced four zones, ZN, ZC, ZE and ZD, which the MS perceives by measuring the RSS it receives from the SBS. The UE makes the measurement request to the SeNB just after entering the ZC, which is earlier than when the signal falls below the usual threshold level commonly set. The SeNB utilizes this lead time for gathering the CL information about all NeNBs from the backbone network and assigns $S_{CL}$ to them. Also, one or more poorly scoring NeNBs are omitted from measurement activities. Since the SeNB completes all the tasks efficiently (in less than fraction of a second) and well within the lead time, practically no HO delay is incurred. Finally, as most failure of the LTE or LTE-A HOs occur before the delivery of the SeNB Handover Command to the UE, it implies the late initiation of the HO process [39]. This is due to the delay in performing the measurements by UE and SeNB receiving the measurement report from the UE before finalizing the TeNB and starting the HO initiation. This is of particular concern in case of fast moving UEs. The proposed HO method can overcome the above mentioned problem by reducing measurement time through shortlisting of NeNBs and consequently improves HO reliability.

## 7. Conclusions

A reliable and fast TeNB selection scheme for LTE-A handover (HO) is proposed in this paper. The HO reliability is achieved by employing orientation matching, current load of eNBs and received signal strength values of eNBs for the final selection of the TeNB. The SeNB assigns individual scores to all the NeNBs for each of the three parameters and selects the TeNB based on the highest weighted average score. The proposed HO scheme is faster than the existing methods of LTE-A HO because the UE measures only the NeNBs shortlisted based on $S_{OM}$ and $S_{CL}$. To efficiently manage the entire process of TeNB selection and HO, a novel concept of RSS-power zones is introduced. An important criterion for the usage of zones is that all HO-related activities are completed before the UE enters into the Zone of Doom. To validate the performance of the proposed system, we conducted various simulation experiments by considering five fixed UE movement trajectories, three different mobility models and the reliability of HOs (i.e., percentage of correct HOs). Results obtained have shown that the HO success rate varied with the underlying movement trajectories and the variation in the current load of the different NeNBs. Manhattan model produced the best results with a maximum of 75% correct HOs for fixed inter-eNB placement. Studying the multi-radio access technology handover performance in a 5G heterogeneous environment involving LTE-A and WiFi networks is suggested as an extension to the work presented here.

APPENDIX A - LIST OF ACRONYMS USED IN THE PAPER

| Acronyms | Full Form |
|---|---|
| AAM | Average Angle of Motion |
| CL | Current Load |
| EAE | Expected Angle of Exit |
| eNB | Evolved Node B |
| eNB_ID | Evolved Node B-Identifiers |
| EPC | Evolved Packet Core |
| GAONB | Geographical Angle of the NeNB |
| GPS | Global Positioning System |
| HO | Handover |
| HOM | Handover Margin |
| LMS | Least-Mean Square |
| LTE-A | Long Term Evolution-Advanced |
| MAC | Medium Access Control |
| NeNB | Neighbouring Evolved Node B |
| OM | Orientation Matching |
| PCT | Polar Coordinate Table |
| PDN | Packet Data Network |
| PGW | Packet Data Network Gateway |
| PTeNB | Potential Target Evolved Node B |
| QAM | Quadrature Amplitude Modulation |
| QPSK | Quadrature Phase-Shift Keying |
| QoS | Quality of Service |
| RAD | Relative Angular Distance |
| RAD_COMPL | RAD Complement |
| RAD_REF | RAD Reference |
| REQ | Request |
| RSS | Received Signal Strength |
| TeNB | Target Evolved Node B |
| TTT | Time-to-Trigger |
| UE | User Equipment |
| VeNBL | Visited Evolved Node B List |
| V2I | Vehicle-to-Infrastructure |
| WiFi | Wireless Fidelity |
| WiMAX | Worldwide Interoperability for Microwave Access |
| ZC | Zone of Concern |
| ZD | Zone of Doom |
| ZE | Zone of Emergency |
| ZN | Zone of Normalcy |
| 3GPP | Third Generation Partnership Project |

| | | |
|---|---|---|
| 4G | | Fourth Generation |
| 5G | | Fifth Generation |